\documentclass[mathptm]{statsoc}
\usepackage[T1]{fontenc}
\usepackage[utf8]{inputenc}
\usepackage{microtype}
\usepackage[colorlinks=false]{hyperref}
\usepackage{amsmath}
\usepackage{amssymb}
\usepackage{commath}

\usepackage[letterpaper]{geometry}

\usepackage{tikz}
\usetikzlibrary{arrows}

\usepackage{booktabs}

\DeclareMathOperator{\E}{\mathbb{E}}
\DeclareMathOperator{\R}{\mathbb{R}}
\DeclareMathOperator{\ind}{\mathbb{I}}
\newcommand{\convd}{\rightsquigarrow}

\title{Self-exciting point processes with spatial covariates: modeling the
  dynamics of crime}

\author[A. Reinhart and J. Greenhouse]{Alex Reinhart and Joel Greenhouse}

\address{Carnegie Mellon University, Pittsburgh, PA}

\email{areinhar@stat.cmu.edu}
\received{August 11, 2017}
\revised{January 24, 2018}
\accepted{February 14, 2018}

\begin{document}
\begin{abstract}
  Crime has both varying patterns in space, related to features of the
  environment, economy, and policing, and patterns in time arising from criminal
  behavior, such as retaliation. Serious crimes may also be presaged by minor
  crimes of disorder. We demonstrate that these spatial and temporal patterns
  are generally confounded, requiring analyses to take both into account, and
  propose a spatio-temporal self-exciting point process model that incorporates
  spatial features, near-repeat and retaliation effects, and triggering. We
  develop inference methods and diagnostic tools, such as residual maps, for
  this model, and through extensive simulation and crime data obtained from
  Pittsburgh, Pennsylvania, demonstrate its properties and usefulness.
\end{abstract}
\keywords{self-exciting point processes; predictive policing; residual maps}

% TODO: Be consistent about \lambda vs. \hat \lambda

\section{Introduction}

\footnotetext{This is the peer reviewed version of the following
    article: A. Reinhart and J. B. Greenhouse, ``Self-exciting point processes
    with spatial covariates: modelling the dynamics of crime,'' \textit{Journal
      of the Royal Statistical Society: Series C (Applied Statistics)}, vol.\
    67, pp.\ 1305--1329, Nov 2018, which has been published in final form at
    \url{https://doi.org/10.1111/rssc.12277}. This article may be used for
    non-commercial purposes in accordance with Wiley Terms and Conditions for
    Use of Self-Archived Versions.}

As police departments have moved to centralized computer databases of crime
reports, models to predict the risk of future crime across space and time have
become widely used. Police departments have used predictive methods to target
interventions aimed at reducing property crime \citep{Hunt:2014vr,Mohler:2015iy}
and violent crime \citep{Ratcliffe:2011dn,Taylor:2011ft}, and to analyze
hotspots of robbery \citep{VanPatten:2009ul} and shootings
\citep{Kennedy:2010ec}, among many other applications. Predictive policing
methods are now widely deployed, with law enforcement agencies routinely making
operational decisions based on them \citep{Perry:2013vh}, and meta-analyses have
shown that these policing programs can result in statistically significant crime
decreases \citep{Braga:2014by}.

Predictive models of crime come in several forms. The most common are tools to
identify ``hotspots,'' small regions with elevated crime rates, using methods
like kernel density estimation or hierarchical clustering on the locations of
individual crimes \citep{Levine:2015}. These tools produce static hotspot maps
that can be used to direct police patrols. A substantial research literature
demonstrates that crime is highly clustered, justifying hotspot methods that
identify clusters for intervention \citep{Braga:2014by,Andresen:2016kt}, though
these methods typically do not model changes in hotspots over time, even though
research suggests that some hotspots emerge and disappear over weeks or months
\citep{Gorr:2015jd}.

Other analysis focuses on ``near-repeats'': a locally elevated risk of crime
immediately after a location experiences a crime, with the risk decaying back to
the baseline level over a period of weeks or months. Near-repeats are often
analyzed using methods borrowed from epidemiology that assess space-time
clustering, such as Knox tests \citep{Ratcliffe:2008hs,Haberman:2012iz}, though
these methods are not very fine-grained, giving only a sense of the distance and
time over which near-repeat effects are statistically significant but not the
form of their decay or uncertainty in their effect. Nonetheless, near-repeat
behavior has been observed for burglaries, possibly because burglars return to
areas with which they are familiar \citep{Townsley:2003bu,Bernasco:2015fl}, and
also with other types of crime, perhaps connected to gang activities and
retaliation attacks \citep{Youstin:2011dl}.

A range of regression-based analyses are also used to predict crime risks. One
approach uses the incidence of ``leading indicator'' offenses as covariates to
predict more serious crimes at later times, and taking leading indicators into
account can improve predictions of crime \citep{Cohen:2007iw,Gorr:2009ed}.
Leading indicators include various minor crimes, such as criminal mischief or
liquor law violations, and police agencies can target intervention if they know
which leading indicators predict which types of crime. On a larger scale, the
``broken windows'' theory states that low-level offenses, if not adequately
controlled, lead to more serious crimes as social control disintegrates
\citep{Kelling:1982}. Research on the broken windows hypothesis has had mixed
results, suggesting the need for further tests of its predictive power
\citep{Harcourt:2006vd,Cerda:2009ex}.

Finally, regression is also used to assess local risk factors for crime. Risk
Terrain Modeling \citep{Kennedy:2010ec,Kennedy:2015gp} divides the city into a
grid, regressing the number of crimes recorded in each grid cell against the
presence of selected risk factors, such as gang territories, bars, high-risk
housing complexes, recent parolees, and so on. The regression output gives
police a quantitative assessment of the ``risk terrain'', and enables directed
interventions targeted at specific risk factors, which can more efficiently use
police resources to reduce crime. The identification of risk factors is also
important for developing criminological theory to understand the nature and
causes of crime \citep{Brantingham:1991}.

Together, these lines of research show the range of statistical methods used to
answer important policing policy questions using historical crime data. In this
paper, we introduce a single self-exciting point process model of crime that
unifies features of all of these methods, accounting for near-repeats, leading
indicators, and spatial risk factors in a single model, and producing dynamic
hotspot maps that account for change over time. We develop a range of
diagnostic and simulation tools for this model. Furthermore, we demonstrate a
serious flaw in previous statistical methods: if leading indicators,
near-repeats, and spatial features are not modeled jointly, their effects are
generically confounded. This confounding may have affected previously published
results. Additional simulations illustrate confounding issues that remain when
some covariates are unmeasured or unknown, making it inherently difficult to
interpret any spatio-temporal model.

The point process model of crime proposed here extends a model introduced by
\citet{Mohler:2011ft} and refined by \citet{Mohler:2014iu}. This model accounts
for changing hotspots and near-repeats by assuming that every crime induces a
locally higher risk of crime that decays exponentially in time; hotspots, where
many crimes occur in a short period of time, decay away unless sustained
criminal activity keeps the crime intensity high. In addition, the model
includes a fixed background to account for chronic hotspots, and allows leading
indicator crimes to contribute to the crime intensity, with weights varying by
crime type and fit by maximum likelihood. \citet{Mohler:2015iy} demonstrated
that a simplified version of this model, used to assign daily patrol priorities
for a large urban police department, can beat predictions by experienced crime
analysts, leading to a roughly 7.4\% reduction in targeted crimes.

We extend the model proposed by \citet{Mohler:2014iu} to incorporate spatial
features, enabling tests of criminological theory; by introducing parameter
inference tools, allowing quantification of near-repeats and tests of leading
indicator parameters; and with residual analysis methods, providing fine-grained
analysis of model fit. The utility of the model is then demonstrated on a large
dataset of crime from Pittsburgh, Pennsylvania. We begin by considering the
confounding factors that make a full spatio-temporal model necessary.

\section{Heterogeneity and state dependence}
\label{heterogeneity}

The risk of crime varies in space and time both because of spatial
heterogeneity---local risk factors for crime, differing socioeconomic status,
zoning, property development, policing patterns, local businesses, and so
on---and through dependence on recent state, such as recent crimes that may
trigger retaliation or signal the presence of a repeat offender. In the
criminological literature, these effects have often been studied separately, but
this is problematic. A long line of research suggests that, in general,
the effects of heterogeneity and state dependence are difficult to distinguish
in observational data and can be confounded \citep{Heckman:1991}. We investigate
this possibility in this section, demonstrating the need for crime models that
control for both effects.

% TODO Is there any review discussing heterogeneity, state dependence, and
% confounding more generally than Heckman's short paper?

Spatial heterogeneity is usually studied with tools like Risk Terrain Modeling
\citep{Kennedy:2010ec,Kennedy:2015gp}, discussed above. At the same time, a
separate line of research has focused on near-repeat and flare-up effects, which
cause short, local bursts of crime activity, with high risks stimulated by
recent criminal activity. Some crimes may occur not because of features of the
local environment but in response to recent crimes in the same area. As
\citet{Johnson_2008} pointed out, however, these two effects may be confounded.
If a particular neighborhood is ``flagged''---that is, has a risk factor that
makes it more attractive to criminals---it will experience a higher rate of
crime, and after any particular crime, the local risk of a repeat offense will
appear to be higher than in other parts of the city without the risk factor. But
this is because of the local risk factor, not because the occurrence of one
crime ``boosted'' the risk temporarily. Boosting and flagging are two
substantively different causal theories of crime, and suggest different policies
and interventions to address their causes, but may be difficult to distinguish
from recorded crime data alone.

To distinguish between these causes, \citet{Johnson_2008} proposed a simulation
approach. A virtual set of households was created, each with a baseline risk of
burglary that depended on separate risk factors, and in each interval of time,
burglaries were simulated based on the risk factors. Separate simulations were
run with and without a boosting effect. Repeated across many simulations, this
produced patterns of near-repeats that could be compared against observed crime
data, and it was found that the simulations containing the boost effect matched
the observed data much better than those without. \citet{Short:2009cf}, in a
similar approach, specified several different stochastic models of crime, and
found that a model incorporating near-repeat behavior fit the observed
distribution of burglaries in Los Angeles much better than one without.

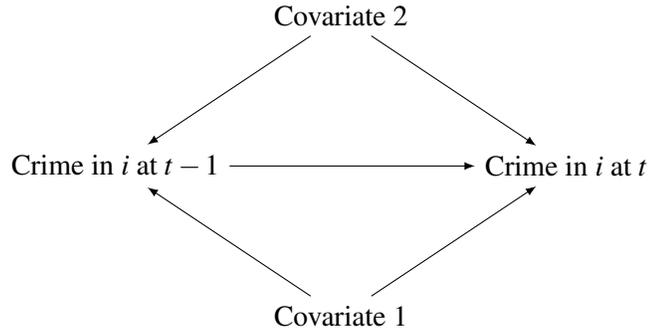
\begin{figure}
  \centering
  \begin{tikzpicture}[>=latex]
    \node (cti) at (6, 0) {Crime in $i$ at $t$};
    \node (cti1) at (0, 0) {Crime in $i$ at $t - 1$};
    \node (cov1) at (3, -2) {Covariate $1$};
    \node (cov2) at (3, 2) {Covariate $2$};

    \draw[->] (cov1) -- (cti);
    \draw[->] (cov1) -- (cti1);
    \draw[->] (cti1) -- (cti);
    \draw[->] (cov2) -- (cti);
    \draw[->] (cov2) -- (cti1);
  \end{tikzpicture}
  \caption{A simplified causal diagram of crime observed in a grid cell \(i\) at
    two times, \(t\) and \(t - 1\), when there are two covariates that may
    affect the rate of crime.}
  \label{confounding-causal}
\end{figure}

However, the ability to distinguish specific models or simulations does not
imply that the two effects are not confounded in general.
Fig.~\ref{confounding-causal} gives a simplified causal diagram
\citep{Pearl:2009co} of near-repeat behavior in one particular grid cell \(i\).
The past occurrence of crime at time \(t-1\) may influence the rate of crime at
\(t\) (the boost effect), shown by an arrow between the two, as may two separate
risk factors, which affect the occurrence of crime at both time points (the
flagging effect). Crucially, if the boost effect is ignored, the flagging effect
of the covariates at time \(t\) is confounded, and vice versa. It may be
possible in specific simulations or in specific stochastic models to distinguish
situations with boosting from those without, but in general, estimates of the
size of each effect will be confounded; to understand spatial risk factors we
must account for the boosting, and to understand boosting we must account for
the spatial risk factors.

\begin{figure}
  \centering
  \includegraphics[width=0.7\textwidth]{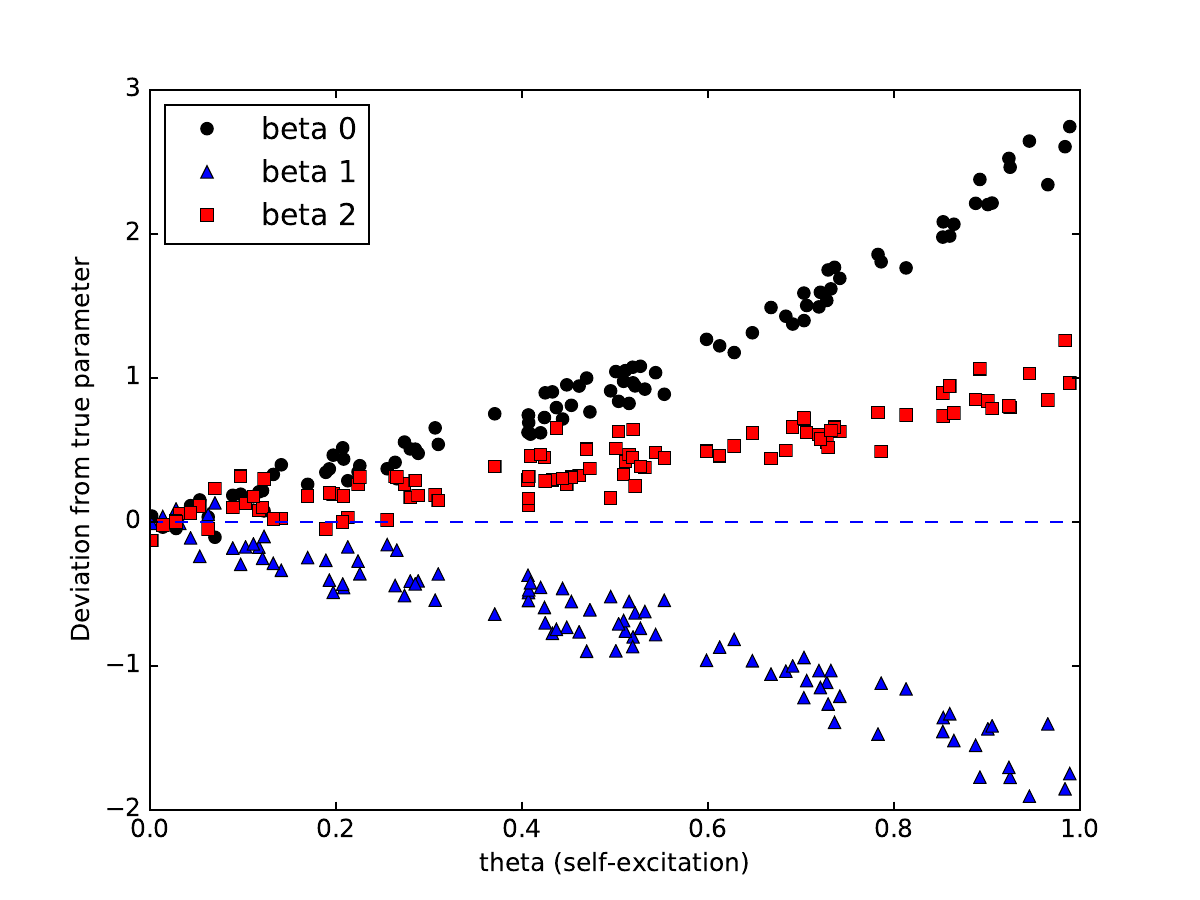}
  \caption{As the near-repeat effect increases from 0 crimes triggered to 1
    crime triggered for every observed crime, spatial Poisson regression
    coefficients gradually become more and more biased.}
  \label{regress-bias}
\end{figure}

A simple simulation can demonstrate this effect. Using the model to be
introduced in Section~\ref{sepp}, we simulate crimes occurring on a grid with
two spatially-varying risk factors for crime, along with a near-repeat effect.
This effect is controlled by a parameter \(\theta\), which specifies the average
number of crimes triggered by each occurring crime. We then perform a spatial
Poisson regression, counting the simulated crimes that occurred in each grid
cell and regressing against the simulated risk factors. The coefficients
\(\beta\) for the intercept and risk factors are shown in
Fig.~\ref{regress-bias}, for simulations ranging from no near-repeat behavior
(\(\theta = 0\)) to a great deal of near-repeats (\(\theta \approx 1\)). As
near-repeats increase, regression coefficients gradually get more biased. The
intercept, \(\beta_0\), increases to account for the additional crimes; the
covariate coefficient \(\beta_1\) decreases from its true value of \(4.8\), and
\(\beta_2\) increases from its true value of \(-2.3\). Notably, both covariate
coefficients shrink towards zero in the presence of near repeats, and the
magnitude of this effect is large compared to their absolute size.

\begin{figure}
  \centering
  \includegraphics[width=2cm]{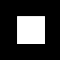}
  \includegraphics[width=2cm]{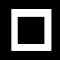}
  \caption{Two synthetic covariates. The covariates have value 1 in the white
    areas and zero elsewhere. The covariate on the left has a true coefficient
    of zero in the simulations, while the covariate on the right has a positive
    true effect. The spatial decay distance is \(\sigma = 5\) pixels, so the
    effect of the right covariate spreads to the area of the left covariate.}
  \label{covariate-images}
\end{figure}

\begin{figure}
  \centering
  \includegraphics[width=0.7\textwidth]{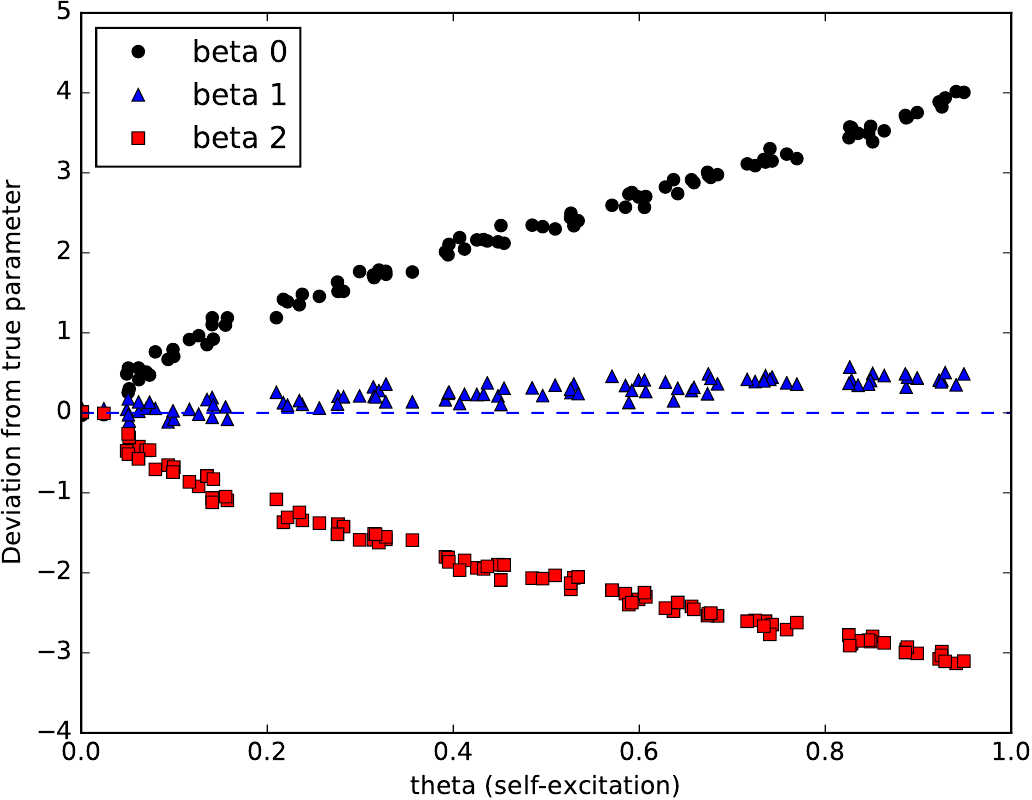}
  \caption{As the amount of self-excitation increases, the coefficient
    \(\beta_1\) (the left covariate in Fig.~\ref{covariate-images}) increases
    from zero, despite its true value being zero. \(\beta_2\) shrinks toward
    zero for the same reason as in Fig.~\ref{regress-bias}.}
  \label{regress-false-positive}
\end{figure}

In certain circumstances, using spatial risk factors with particular patterns,
near-repeats can cause false positives: risk factors that appear related to
crime rates but are not. For example, Fig.~\ref{covariate-images} shows two
synthetic spatial covariates. One is nonzero in a center square, the other in a
ring around that square. Only the first covariate has a true nonzero
coefficient, but because the near-repeat effect produces crimes slightly outside
the square, its effect ``leaks'' to the outer ring, causing the second covariate
to appear to have a positive coefficient, as shown in the simulation results in
Fig.~\ref{regress-false-positive}.

% TODO: Talk about lags?

It is hence clear that methods to estimate spatial risk factors must take into
account near repeats or suffer bias and potentially false positives in their
estimated coefficients. In Section~\ref{omitted-covariates}, further simulations
using the model to be developed below will show the opposite effect:
unaccounted-for spatial risk factors bias estimates of the rate of near repeats,
potentially resulting in estimates that overestimate the boost effect. To
resolve these problems, we propose a self-exciting point process model for crime
that can account for both near repeats and spatial risk factors simultaneously,
eliminating the confounding.

% TODO Maybe mention why Knox with permutation null still works okay?

\section{Methods}

\subsection{Self-exciting point process model}
\label{sepp}

Self-exciting point process models are a class of models for spatio-temporal
point process data that incorporate ``self-excitation'': each event may excite
further events, by locally increasing the event rate for some period of time.
This corresponds to the near repeat phenomenon we need to account for.
Self-exciting point processes are a development of Hawkes processes
\citep{Hawkes:1971}, which are purely temporal processes. The theory and
applications of self-exciting spatio-temporal point processes were reviewed by
\citet{Reinhart:2017}; we give only a brief summary here.

A spatio-temporal point process is characterized by its conditional intensity
function, defined for locations \(s \in X \subseteq \R^d\) and times \(t \in [0,
T)\) as
\begin{equation}\label{intensity-expectation}
  \lambda(s, t \mid {\cal H}_t) = \lim_{\Delta s, \Delta t \downarrow 0} \frac{\E\left[ N\left(B(s,
        \Delta s)  \times [t, t + \Delta t)\right) \mid {\cal
        H}_t\right]}{|B(s, \Delta s)| \Delta t},
\end{equation}
where \(|B(s, \Delta s)|\) is the Lebesgue measure of the ball \(B(s, \Delta
s)\) with radius \(\Delta s\), \(N(A)\) is the counting measure of events over
the set \(A \subseteq X \times [0,T)\), and \({\cal H}_t\) is the history of
events in the process up to time \(t\). In the limit, the conditional intensity
can be interpreted as the instantaneous rate of events per unit time and area,
and hence the expected number of events in time interval \([t_1, t_2)\) and
region \(B\) is
\[
  \E[N(B \times [t_1, t_2))] = \int_{t_1}^{t_2} \int_B \lambda(s, t) \dif s \dif t.
\]
Self-exciting point processes have conditional intensities of the form
\begin{align*}
  \lambda(s, t \mid {\cal H}_t) &= \mu(s) + \sum_{i: t_i < t} g(s - s_i, t -
                                  t_i)\\
  &= \mu(s) + \int_0^t \int_X g(s - u, t - r) \dif N(u \times r),
\end{align*}
where \(g\) is a triggering function that determines the form of the
self-excitation. In the remainder of this paper, we will refer to \(\lambda(s,
t)\) without the explicit \({\cal H}_t\) to simplify notation, but \(\lambda(s,
t)\) should still be understood to depend on the past history of the process.

Models of this form have been widely used in a variety of processes exhibiting
clustering, such as earthquake epicenters \citep{Ogata:1999jz} and the
occurrence of infectious diseases \citep{Meyer:2011ct,Meyer:2014in}.
\citet{Mohler:2011ft} developed such a model for the occurrence of violent crime
by building on the models used in earthquake forecasting, known in the
seismology literature as epidemic-type aftershock sequence models
\citep{Ogata:1999jz}. This model allows hotspot estimates to change over time by
separating crime into chronic hotspots, which remain fixed in time, and
temporary hotspots, which are caused by increases or changes in crime. (In
seismological models, earthquakes are similarly divided into main shocks and
aftershocks triggered by those main shocks.) Hotspot intensities are modeled
with a modification of kernel density smoothing, where past crimes contribute to
the intensity with effects that decay away in time, and the bandwidth parameters
are estimated to best fit the data instead of being chosen by the operator.

\citet{Mohler:2014iu} further adapted the model to include leading indicator
crimes, producing a model that predicts the conditional intensity \(\lambda(s,
t)\) of crime at each location \(s\) and time \(t\) as the sum of a background
rate and a sum of functions of prior crimes:
\begin{equation}\label{mohler-intensity}
  \lambda(s, t) = \mu(s) + \sum_{\substack{\text{all events } i\\\text{
        before time } t}} g(s - s_i, t - t_i, M_i),
\end{equation}
where \(\mu(s)\) is a background crime rate that does not vary in time, and
\(s_i\) and \(t_i\) are locations and times of other crimes used as leading
indicators. \(M_i\) represents the \textit{type} of each leading indicator
crime, as different indicators are allowed to have different predictive effects
in this model.

The triggering function \(g\) is defined to be
\[
  g(s, t, M) = \frac{\theta_M \omega}{2 \pi \sigma^2} \exp(-\omega t) \exp\left(
    - \frac{\|s\|^2}{2 \sigma^2}\right),
\]
where \(\sigma^2\) is the bandwidth, \(\theta_M\) determines how much each type
of leading indicator contributes to the intensity, and the effect decays
exponentially in time with a rate controlled by \(\omega\). Because \(g\) is
chosen to integrate to \(\theta_M\), it has a natural interpretation: the
expected number of target crimes induced by a single leading indicator crime of
type \(M\).

\citet{Mohler:2014iu} chose the background crime rate \(\mu(s)\) to be a sum of
weighted Gaussian kernels centered at prior crimes:
\begin{equation}\label{mohler-bg}
  \mu(s) = \sum_i \frac{\alpha_{M_i}}{2 \pi \eta^2 T} \exp\left( -
    \frac{\|s\|^2}{2 \eta^2}\right).
\end{equation}
Here \(\alpha_{M_i}\) determines the contribution of each leading indicator type
to the background rate, \(\eta^2\) is the bandwidth, and \(T\) is the total
length of time over which the crime data falls.

This model has several limitations. The background component \eqref{mohler-bg}
does not explicitly account for varying spatial features or give estimates of
their effects, and the use of weighted Gaussian kernels for both \(g(s, t, M)\)
and \(\mu(s)\) makes the model parameters difficult to identify; to prevent
multiple modes in the log-likelihood, \citet{Mohler:2014iu} had to set \(\sigma
= \eta\).

We have extended this model to replace the nonparametric \(\mu(s)\) with one
that directly incorporates spatial covariate information, allowing estimates of
the effects of each covariate and avoiding identifiability issues. We assume
that the observation domain \(X\) is divided into cells \(c\) of arbitrary
shape, inside of which a covariate vector \(X_c\) (including an intercept term)
is known, resulting in the model
\begin{equation}\label{revised-model}
  \lambda(s, t) = \exp\left( \beta X_{C(s)} \right) + \sum_{i : t_i < t} g(s -
  s_i, t - t_i, M_i),
\end{equation}
where \(C(s)\) is the index of the covariate cell containing \(s\) and the
triggering function \(g\) is unchanged. We let \(g(s, t, M) = 0\) for \(s <
\delta\), for an arbitrary short distance \(\delta\), to prevent crimes that
occur at exactly the same location from enticing the model to converge to
\(\sigma = 0\).

In principle, this model could be built with covariates that vary continuously
in space, defined by a function \(X(s)\). This would increase the generality of
the model. However, in practice, this generality is not necessary: most
socioeconomic, demographic, or land use variables are observed only in cells
such as city blocks, census blocks, or neighborhoods. Piecewise constant
covariates also make estimation and simulation more computationally tractable,
and so the small loss in generality is worth the substantial gain in
practicality.

We may also reasonably ask about the form of the triggering function \(g\),
which specifies an exponential decay in time and a Gaussian kernel in space.
\citet{Meyer:2014in}, for example, analyzing the spread of infectious disease,
proposed a power law kernel to account for long-range flows of people.
Unfortunately, most alternate spatial kernels make the expectation maximization
strategy described below more difficult, by making analytical maximization on
each iteration impossible. These kernels could still be used, but with the
additional computational cost of numerical maximization.

\subsection{Simulation algorithm}
\label{simulation-algorithm}

Simulations from self-exciting point process models have proved useful both for
examining inference and for the simulation studies discussed in
Sections~\ref{heterogeneity} and \ref{simulation-studies}. Various simulation
algorithms for self-exciting point processes were discussed by
\citet[Section~3.3]{Reinhart:2017}. We chose to use an algorithm introduced by
\citet{Zhuang:2004ex} for earthquake aftershock sequence models, which is fast
and efficient for our model structure. This algorithm draws on a key property of
self-exciting point processes shown by \citet{Hawkes:1974ib}: a self-exciting
process can be represented as a cluster process. Cluster centers come from an
inhomogeneous Poisson process with rate \(\mu(s)\), and each cluster center
produces a cluster of offspring events with locations and times determined by
the triggering function \(g(s, t, M)\). Each of these offspring events may
trigger further offspring of its own, and so on.

This leads to a natural simulation procedure that first draws from the
inhomogeneous Poisson cluster center process, then draws a generation of
offspring based on those cluster centers, and repeats until there are no more
offspring. Full details are given by \citet{Zhuang:2004ex}. Draws from the
cluster center process are made easier by our assumption that the observation
domain is divided into cells \(c\), inside each of which is a constant covariate
vector \(X_c\); we can hence draw cluster centers from homogeneous Poisson
processes inside each cell.

Our simulation system can simulate from the model specified by
eq.~\eqref{revised-model}, but can also simulate various violations of
assumptions: the spatial distribution of offspring can be Gaussian, \(t\) with
arbitrary degrees of freedom, Cauchy, or various other shapes, and their
temporal distribution can be drawn from an exponential distribution or a Gamma
distribution with arbitrary parameters. The framework is flexible and allows
additional distributions to be chosen easily; this feature will be used in
Section~\ref{misspecification} to test model performance under various types of
misspecification.

\subsection{Parameter inference}
\label{parameter-inference}

\citet{Mohler:2014iu} fit the self-exciting model by maximum likelihood, using
the log-likelihood function for spatio-temporal point processes:
\begin{equation}\label{loglik}
  \ell(\Theta) = \sum_i \log \lambda(s_i, t_i) - \int_0^T \int_X
  \lambda(s, t) \dif s \dif t,
\end{equation}
where \(X\) is the spatial domain of the observations and \(\Theta\) a complete
vector of parameters. The log-likelihood is optimized via expectation
maximization, using the approximation that \(X = \R^2\) to simplify calculation
of the triple integral, which is valid when most crime triggered by the observed
crimes occurs within the study area \citep{Schoenberg:2013}. Note that, as the
model predicts the target crime and not the leading indicators, only the target
crimes are included in the sum in Eq.~\eqref{loglik}. We adapted the expectation
maximization procedure to fit our extended model.

The expectation maximization procedure for self-exciting processes was first
described by \citet{Veen:2008cr}, and follows the general procedure described by
\citet[Section 3.1]{Reinhart:2017}. A latent variable \(u_i\) is introduced for
each event \(i\), indicating whether the event came from the background process
(\(u_i = 0\)) or was triggered by a previous event \(j\) (\(u_i = j\)).
Augmented with this variable, the log-likelihood simplifies from the form in
\eqref{loglik}, since the intensity \(\lambda(s_i, t_i)\) no longer involves a
sum over the background \(\mu(s)\) and all previous events but only the term
indicated by \(u_i\). We can then take the expectation and maximize. In our
model, maximization proceeds in closed form for most parameters, apart from
\(\beta\), which must be separately numerically maximized on each iteration.

\citet{Mohler:2014iu} did not provide inference for the self-exciting point
process model parameters, though they may be of interest: the self-excitation
parameters \(\sigma^2\), \(\omega\), and \(\theta\) may be used to test
hypotheses about the concentration of crime and the nature of leading
indicators. Mohler's nonparametric background \eqref{mohler-bg} also does not
incorporate spatial covariates, though \(\beta\) in our background indicates the
association between spatial covariates and crime. There are several potential
routes to deriving asymptotic confidence intervals for these parameters, which
we consider in turn.

First, \citet{Rathbun:1996bf} demonstrated the asymptotic normality and
consistency of the maximum likelihood estimator for spatiotemporal point
processes: \(\sqrt{T}(\hat \Theta - \Theta) \convd_u N(0, \Sigma)\) as \(T \to
\infty\), where \(\convd_u\) represents uniform convergence in distribution and
\(\Theta\) is the complete vector of parameters. The result holds under certain
regularity conditions on the conditional intensity. \citet{Rathbun:1996bf}
suggested an estimator for the covariance matrix \(\hat \Sigma\) of the
parameter estimate \(\hat \Theta\),
\[
  \hat V(\hat \Theta) = \left( \sum_{i=1}^{N_c} \frac{\Delta(s_i,
      t_i)}{\lambda(s_i, t_i)}\right)^{-1},
\]
where \(\Delta(s,t)\) is a matrix-valued function with elements
\begin{equation}\label{deltafunc}
  \Delta_{ij}(s, t) = \frac{\pd{}{\hat \Theta_i} \lambda(s,
    t) \pd{}{\hat \Theta_j} \lambda(s, t)}{\lambda(s, t)}.
\end{equation}
With the full estimated covariance matrix, we calculated standard errors for
each estimator, and produced confidence intervals from these.

An alternate approach, following again from the asymptotic normality result, is
to use the observed information matrix at the maximum likelihood estimate, based
on the Hessian of the log-likelihood:
\[
  \hat \Sigma = - H(\hat \Theta)^{-1},
\]
where \(H(\hat \Theta)\) is the matrix of second partial derivatives of
\(\ell(\Theta)\) evaluated at \(\hat \Theta\). This approach was suggested by
\citet{Ogata:1978dt} in the context of an asymptotic normality result for
temporal point processes, and has been frequently used for spatio-temporal
models in seismology; however, \citet{Wang:2010dw}, comparing its estimated
standard errors with those found by repeated simulation, found that it can be
heavily biased for smaller samples.

Nonetheless, we implemented the estimator using Theano \citep{Bergstra:2010}, a
Python package for describing computations that automatically generates fast C
code and automatically computes all necessary derivatives. We then performed a
series of 350 simulations to compare the finite-sample performance of both
estimators with our model, using randomly chosen parameter values, obtaining the
results shown in Table~\ref{coverage-comparison}. Coverage is worst for the
self-excitation parameters \(\sigma^2\), \(\omega\), and \(\theta\), which are
affected by any remaining boundary effect (see Section~\ref{boundary-effects})
not compensated for by the buffer region; Rathbun's covariance estimator
achieves nearly nominal coverage for \(\beta\), which is less affected. Overall,
Rathbun's estimator achieves 88\% coverage and is closest to its nominal 95\%
coverage. We will use this estimator in our analysis in Section~\ref{results}.

\begin{table}
  \caption{\label{coverage-comparison}Coverage of nominal 95\% CIs}
  \centering
  \begin{tabular}{l r r}\toprule
    \textbf{Variable} & \textbf{Hessian (\%)} & \textbf{Rathbun (\%)} \\ \midrule
    $\sigma^2$ & 86 & 88 \\
    $\omega$ & 87 & 91 \\
    $\theta$ & 82 & 63 \\
    $\beta_0$ & 77 & 83 \\
    $\beta_1$ & 89 & 92 \\
    $\beta_2$ & 86 & 89 \\
    \midrule
    \textbf{Average} & 85 & 88 \\
    \bottomrule
  \end{tabular}
\end{table}

\subsection{Residual analysis}
\label{residual-maps}

Once a model is fit, it is useful to be able to determine \emph{where} the model
fits: what types of systematic deviations are present, where covariates may be
lacking, what types of crimes are over- or under-predicted, and so on.
Eq.~\eqref{intensity-expectation} suggests we can produce these detailed
analyses: because the point process model predicts a conditional intensity at
each location, we can calculate the expected number of crimes within each region
in a certain period of time, and compare this against the true occurrences over
the same time, producing a residual map. These residuals are defined to be
\citep[chapter 15]{Daley:2008v2}
\[
  R(h) = \int_{\R \times \R^2} h(s, t) \left[ N(\dif s \times \dif t) -
    \lambda(s, t) \dif t \dif s \right],
\]
where \(N(\cdot)\) is the counting measure of events in the given region, and
\(h(s, t)\) is a bounded window function. Typically, \(h(s, t)\) is taken to be
an indicator function for a chosen spatio-temporal region.

To calculate \(R(h)\), a typical approach is to choose a time window \([t_1,
t_2)\)---say, a particular week or month---and integrate the conditional
intensity over this window, producing an integrated intensity function
\[
  \lambda(s) = \int_{t_1}^{t_2} \lambda(s, t) \dif t.
\]
Then the spatial region \(X\) is divided appropriately and the intensity is
integrated over each subdivision, then compared against the number of events in
that subdivision during that time window.

Choosing spatial subdivisions for residuals requires care. The obvious choice is
a discrete grid, but the right size is elusive: small grid cells produce skewed
residuals with high variance (as most cells have no crimes), and positive and
negative residual values can cancel each other out in large cells.
\citet{Bray:2014eo} instead suggest instead using the Voronoi tessellation of
the plane, which produces a set of convex polygons, known as Voronoi cells, each
of which contains exactly one crime and all locations that are closer to that
crime than to any other.

Given this tessellation, the raw Voronoi residuals \(\hat r_i\) for each cell
\(C_i\) are
\[
  \hat r_i = 1 - \int_{C_i} \hat \lambda(s) \dif s.
\]
The choice of Voronoi cells ensures that cell sizes adapt to the distribution of
the data, and \citet{Bray:2014eo} cite extensive simulations by
\citet{Tanemura:2003sw} indicating that the Voronoi residuals of a homogeneous
Poisson process have an approximate distribution given by
\[
  \hat r_i \sim 1 - X, \qquad \text{where } X \sim \text{Gamma}(3.569, 3.569),
\]
so that \(\E[\hat r_i] = 0\). (Here the gamma distribution is parametrized by
its shape and rate.) But because the conditional intensity function
\eqref{revised-model} is not homogeneous, we performed similar simulations for
random parameter values, fitting to 1,332,546 simulated residuals by maximum
likelihood the approximate distribution \(X \sim \text{Gamma}(3.389, 3.400)\).

After each \(\hat r_i\) is found, using Monte Carlo integration over \(C_i\),
the Voronoi cells can be mapped with colors corresponding to their residual
values. To ease interpretation, colors are determined by \(-\Phi^{-1}(F(1 - \hat
r_i))\) where \(F\) is the cumulative distribution function of the approximate
distribution of \(X\) and \(\Phi^{-1}\) the inverse normal cdf. Positive
residuals hence indicate more observed crime than was predicted, and negative
residuals less.

These residual maps provide much more detailed information than previous global
measures of hotspot fit, and can indicate areas with unusual patterns of
criminal activity. For example, consider a model that predicts homicides using
leading indicators such as assault and robbery; this model may perform well in
an area that experiences gang-related violence, but would systematically
over-predict homicides in a commercial area full of bars and nightclubs, where
most assaults are drunken arguments rather than signs of gang conflict. An
example residual map is given in Fig.~\ref{burglary-resids}
(Section~\ref{burglary-analysis}) for Pittsburgh burglary data, illustrating the
use of this method.

The example map does illustrate one weakness of Voronoi residual maps. We would
expect areas with large positive residuals (red, in the map) to have a higher
crime density than areas with large negative residuals (blue), since positive
residuals indicate more crimes occurred than were expected. Hence areas with
positive residuals tend to have smaller Voronoi cells than areas with negative
residuals, and the map is visually dominated by large cells with negative
residuals. Closer inspection reveals clusters of very small cells containing
large positive residuals; these are the locations of new crime hotspots. Users
should be aware of this problem when interpreting residual maps.

We have also introduced animated residual videos. Instead of a single time
window \([t_1, t_2)\), we produce a succession of windows \(\{[t_1 + (i - 1)
\delta_t, t_1 + i \delta_t) : i = 1,2,3\dots\}\). For each window, we calculate
the Voronoi tessellation of crimes occurring in that window and the
corresponding residuals \(\hat r_j\). These residuals, and the times of the
events defining each cell, are used to build a smoothed residual field similar
to that suggested by \citet{Baddeley:2005bc}. The residual value at each
animation frame and each point in space is determined by a kernel smoother,
using an exponential kernel in time and a Gaussian kernel in space, with the
same structure as the triggering function \(g(s, t)\). An animated version of
Fig.~\ref{burglary-resids} is provided in the Supplemental Materials as an
example.

A purely temporal residual analysis can be useful to illustrate the calibration
of the model over time. Consider plotting the index \(i\) of each event versus
the quantity
\[
  \tau_i = \int_0^{t_i} \int_X \lambda(s, t) \dif s \dif t,
\]
the expected number of events in the interval \([0, t_i)\). This is an extension
of the standard transformation property of point processes: if the model is
correct, the resulting process \(\{\tau_i\}\) will be a stationary Poisson
process with intensity 1 \citep{Papangelou:1972qy}. Hence the plotted points
will fall on the diagonal, and by plotting the deviation from the diagonal, poor
calibration becomes obvious. Similar diagnostics have previously been used for
seismological models \citep[e.g.][]{Ogata:1988nr}. An example of this diagnostic
will be shown in Section~\ref{omitted-covariates}, demonstrating its use in
detecting some forms of model misspecification.

\subsection{Prediction evaluation}
\label{evaluation}

To compare different methods for locating crime hotspots, fairly simple metrics
have been typically used, such as the hit rate: the percentage of crimes during
the test period that occur inside the selected hotspots. A modified version is
the Prediction Accuracy Index (PAI), which divides the hit rate by the total
fraction of the map that is selected as hotspots, to penalize methods that
achieve their accuracy by simply selecting a larger total land area
\citep{Chainey:2008dn}. However, this still requires selecting a single set of
hotspots, and in some simulations, we found the PAI was maximized by shrinking
the denominator, selecting a single 100 meter grid cell containing several
crimes as the only hotspot. This is hardly practical, and says little about the
comparative performance of models. The conditional intensity function
\(\lambda(s, t)\) provides much richer information: the estimated rate of crime
at every location at all times. We would like a metric that is maximized when
\(\hat \lambda(s, t)\) neither underestimates nor overestimates the true crime
rate.

Such a metric can be found with proper scoring rules \citep{Gneiting:2007},
which have previously been used for self-exciting point process models in
seismology \citep{VereJones:1998}. Scoring rules evaluate probabilistic
forecasts of events: a score \(S(P, x)\) returns the score of a predictive
distribution \(P\) when outcome \(x\) occurs. A scoring rule \(S\) is
\textit{proper} if the expected value of \(S(P, x)\) is maximized by \(P\) when
\(x\) is drawn from \(P\). An example of a proper score is the logarithmic score
\(S(P, x) = \log p_x\), where \(p_x\) is the forecast probability of event \(x\)
under the predictive distribution \(P\). The expected value of the logarithmic
score, under a particular \(P\), can be interpreted as the predictability of the
outcome \(x\), and is related to the entropy of the distribution.

\citet{VereJones:2005}, noting this connection, proposed a method for comparing
different predictive models. The \emph{relative entropy} of a predictive
distribution \(P\) compared to a baseline distribution \(\pi\) is
\[
  I^* = \E_P \log \frac{p_x}{\pi_x},
\]
where \(\pi\) is a simple default distribution, such as a homogeneous Poisson
process model, against which all models are compared. Applied to a self-exciting
point process model, we may produce \(P\) by performing one-step predictions:
after each event, form a predictive distribution for the next event. Because the
predictive distribution \(P\) is conditional on the past history of the point
process, \(I^*\) is random, depending on the particular realization of the
process; the average over all possible realizations \(G = \E[I^*]\) is called
the expected information gain, and numerically quantifies the intrinsic
predictability of the process.

A further connection soon becomes apparent. If we perform this one-step
prediction process for each event in a point process realization, the
logarithmic score for each event is the log-likelihood of that event, and the
relative entropy \(I^*\) is the expected log-likelihood ratio. The expected
information gain \(G\) is hence estimated by the log-likelihood ratio on an
observed dataset:
\begin{equation}\label{info-gain}
  \hat G = \frac{1}{T} \log L_1 / L_0,
\end{equation}
where \(L_0\) is the baseline model likelihood and \(L_1\) the likelihood of the
model of interest. The likelihood ratio between two models hence estimates the
difference in score between them, in the form of the relative entropy. (The
theoretical aspects here were reviewed in more depth by \citet{Daley:2004}.)
This quantity an be used to compare the predictive performance of models on test
time periods.

Further, this quantity can be connected to the difference in Akaike Information
Criterion (AIC) between the two models \citep{VereJones:2005}. If the baseline
model has \(k_0\) parameters and the model of interest has \(k_1\) parameters,
the difference in AIC can be written as
\[
  \frac{\Delta \text{AIC}}{2T} = \frac{k_1 - k_0}{T} - \hat G,
\]
where \(\Delta \text{AIC} = 2(k_1 - k_0) - 2 \log L_1 / L_0\). This suggests the
use of \(\Delta \text{AIC}\) to compare the predictive performance of models
with varying number of parameters, which will be demonstrated in
Section~\ref{burglary-analysis}.

\section{Simulation studies}
\label{simulation-studies}

\subsection{Boundary effects}
\label{boundary-effects}

As noted by \citet{Zhuang:2004ex} and \citet{Reinhart:2017}, boundary effects
can be a problem if events are only observed in a subset of the space, such as
if crimes are only recorded inside a specific jurisdiction. If crimes are only
observed in the region \(X\) and time interval \([0, T)\), but also occur
outside \(X\) and at \(t < 0\) or \(t \geq T\), maximum likelihood parameter
estimates can be biased by boundary effects. Unobserved crimes just outside
\(X\) or before \(t = 0\) can produce near repeats that are observed, and
observed crimes near the boundary of \(X\) can stimulate near repeats outside
the boundary that are not. This biases model fits to underestimate the rate of
near repeats.

These boundary effects are distinct from boundary effects in kernel density
estimation \citep[e.g.][]{Cowling:1996kd}, which bias density estimates near the
boundary. Similar problems occur here, with \(\lambda(s, t)\) biased near the
boundary of \(X\), but additional biases on parameter estimates occur. The
nature of these boundary effects can be seen clearly from the parameter updates
in the M step of the EM algorithm. For example, the update step for \(\theta_L\)
is
\[
  \theta_L = \frac{\sum_{\text{responses } i} \sum_{t_j < t_i} P(u_i = j)
    \ind(M_j = L)}{K_L - \sum_{\text{crimes } i} \ind(M_i = L) e^{-(T - t_i) /
      \omega}},
\]
which can be interpreted as a weighted average: for all crimes of type \(L\),
sum up their contributions to response crimes (measured by \(P(u_i = j)\)), and
take the average. An average of \(0.5\), for example, says a crime of type \(L\)
can be expected to contribute to about \(0.5\) future response crimes. The
denominator also contains a temporal boundary correction term that is
negligible when \(T\) is very large.

Suppose, however, that many crimes of type \(L\) occur near the boundary of the
observation region \(X\), and trigger response crimes that occur outside of
\(X\). These response crimes will not be included in the sum in the numerator,
and hence \(\theta_L\) will be biased downward. Updates for \(\sigma^2\) and
\(\omega\) can also be interpreted as weighted averages, and are subject to
similar biases.

\citet{Harte_2012} explored the effects of these biases on the seismological
models. One common workaround to reduce the bias is to introduce a region \(X_0
\subset X\), chosen so that events inside \(X_0\) have triggered offspring that
mostly occur within \(X\). All events in \(X\) contribute to the intensity
\(\lambda(s, t)\), but the weighted averages in the M step only average over
events inside \(X_0\): that is, to update \(\theta_L\), we average over events
of type \(L\) within \(X_0\), counting their contributions to any response
crimes within \(X\). Since most of their offspring will be within \(X\) by
construction, the average will not leave much out.

The same subsetting is also done in time, so only events in the interval \([0,
T_0)\) are considered, where \(T_0 < T\). This eliminates bias caused by events
at \(t\) close to \(T\) triggering offspring that occur after \(T\) and are
hence not observed.

Of course, averaging over events only in a subset of space and time reduces the
effective sample size of the fit, introducing additional variance to parameter
estimates. It does, however, dramatically reduce bias. To demonstrate this,
Table~\ref{simulated-fits} shows parameter values obtained from 50 simulations
from a model with known parameter values, with two covariates. The true
parameters are \(\theta = 0.5\), \(\omega = 7\) days, and \(\sigma = 4\) feet;
the covariate coefficients are \(\beta_1 = 1.2\) and \(\beta_2 = -1.5\). The
grid is \(66 \times 60\) feet and no boundary correction was applied, resulting
in the biases shown. Note that \(\hat \theta\) is biased too low, since events
triggered outside the grid were not observed, and both \(\hat \omega\) and
\(\hat \sigma\) are also too small. The covariate coefficients are both biased
towards zero because the intercept increased to account for the events no longer
accounted for by \(\hat \theta\).

The third column of Table~\ref{simulated-fits} shows the average fit obtained
when an 8-foot boundary was established around the images, so \(X_0\) was the
inner \(50 \times 44\) box; the simulated events occurred over the course of two
years, of which the last thirty days were also left out. These fits suffer from
much less bias.

\begin{table}
  \caption{\label{simulated-fits}Average parameter values with and without
    boundary correction}
  \centering
  \begin{tabular}{l r r}\toprule
    \textbf{Parameter} & \textbf{Uncorrected} & \textbf{Corrected} \\ \midrule
    $\theta$ & 0.3367 & 0.4706 \\
    $\omega$ & 6.104 days & 6.638 days \\
    $\sigma^2$ & 3.173 feet & 3.913 feet \\
    $\beta_0$ (intercept) & -19.65 & -19.78 \\
    $\beta_1$ & 1.135 & 1.176 \\
    $\beta_2$ & -1.348 & -1.498 \\
    \bottomrule
  \end{tabular}
\end{table}

\subsection{Model misspecification}
\label{misspecification}

\begin{figure}
  \centering
  \includegraphics[width=0.4\textwidth]{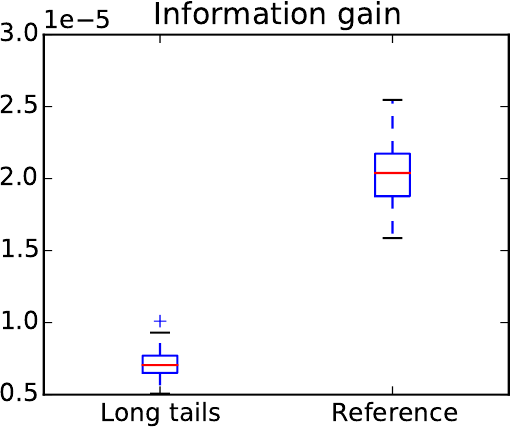}
  \caption{Boxplot of information gains (eq.~\eqref{info-gain}) obtained from
    fits to simulated data with Cauchy-distributed offspring (left) or Gaussian
    offspring (right). The poor fit from model misspecification is noticeable.}
  \label{long-tail-gains}
\end{figure}

In this section we explore the results of model misspecification on the fit, to
determine when misspecification may be detected and corrected. As an example,
consider two simulations: one in which event offspring are drawn from the
Gaussian \(g\) used in fitting our model, and one in which event offspring are
drawn from a Cauchy distribution, giving them a heavy tail that is not
accounted for by our model. Running 100 simulations under each condition and
calculating the log-likelihoods of fits to each, we obtained the information
gains \(\hat G\) shown in Fig.~\ref{long-tail-gains}, which demonstrate the
deterioration in model fit when misspecified. In this situation, the disturbance
in model fit is limited to the self-excitation parameters \(\theta\) and
\(\omega\) (\(\sigma^2\) is not meaningful to compare here), along with the
intercept \(\beta_0\); the estimates of \(\beta\) for the simulated covariates
are unaffected, suggesting that misspecification of the triggering function need
not harm inference about the spatial covariates.

We performed several other simulations of different forms of misspecification,
using boxcar and double exponential spatial distributions and also a Gamma
distribution for offspring times. With spatial misspecification, the covariate
coefficients were still unbiased on average, with slight biases in \(\theta\)
depending on the type of misspecification, and larger biases in \(\omega\)
(towards longer decay times). Misspecification of the offspring timing did not
bias \(\theta\), \(\beta\), or \(\omega\), but did cause systematic
understimation of \(\sigma^2\). These results suggest that the covariate
coefficient estimates of the model are robust to misspecification of the
self-excitation component, though the self-excitation parameters can be
sensitive to misspecification, giving misleading estimates of cluster size and
duration.

\subsection{Unobserved covariates and confounding}
\label{omitted-covariates}

Section~\ref{heterogeneity} discussed the inherent confounding that can occur
when estimating the effect of spatial covariates on crime without accounting for
self-excitation. Fig.~\ref{confounding-causal} demonstrated that this
confounding is generic, occurring whenever there are covariates that affect
crime over time. By building a self-exciting point process model that accounts
for self-excitation and covariates, we can account for both and avoid the
confounding.

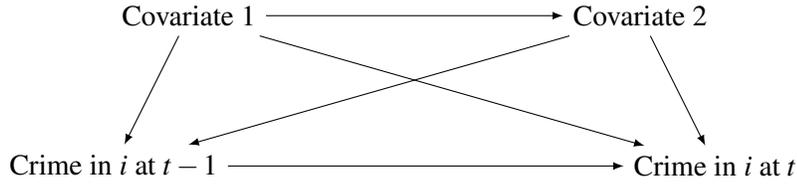
\begin{figure}
  \centering
  \begin{tikzpicture}[>=latex]
    \node (cti) at (8, 0) {Crime in $i$ at $t$};
    \node (cti1) at (0, 0) {Crime in $i$ at $t - 1$};
    \node (cov1) at (1, 2) {Covariate $1$};
    \node (cov2) at (7, 2) {Covariate $2$};

    \draw[->] (cov1) -- (cti);
    \draw[->] (cov1) -- (cti1);
    \draw[->] (cti1) -- (cti);
    \draw[->] (cov2) -- (cti);
    \draw[->] (cov2) -- (cti1);
    \draw[->] (cov1) -- (cov2);
  \end{tikzpicture}
  \caption{A simplified causal diagram depicting potential confounding:
    covariate 1 has a causal relationship with both covariate 2 and crime rates,
    and so if it is unobserved, estimates of covariate 2's effect will be
    confounded.}
  \label{confounding-omitted}
\end{figure}

We must, however, be aware of other types of confounding that can creep in. The
most common is an unobserved covariate: there are many spatial factors that can
influence crime rates, and it is unlikely we can directly measure all of them.
Fig.~\ref{confounding-omitted} demonstrates the danger. A covariate may be
causally related to another covariate as well as to crime rates, and if it is
not observed and accounted for, the other covariate's estimated effect will be
confounded. This is directly analogous to the situation in ordinary regression,
when unobserved predictors may confound regression coefficient estimates.

On the other hand, if the two covariates are \emph{not} correlated in any way,
omitting one does not bias estimates of the other's effect; in traditional
regression its mean effect is simply added to the intercept and the individual
effects simply add to the error variance. However, in the more complicated
self-exciting point process model, omitted covariates may have other detrimental
effects. Though \citet{Schoenberg:2016vy} suggests that parameter estimates
remain asymptotically consistent when covariates are omitted, provided the
effects of those covariates are sufficiently small, a series of simulations
demonstrate the bias that appears in finite samples.

% TODO Write out the Gaussian processes in notation

\begin{figure}
  \centering
  \includegraphics[width=0.8\textwidth]{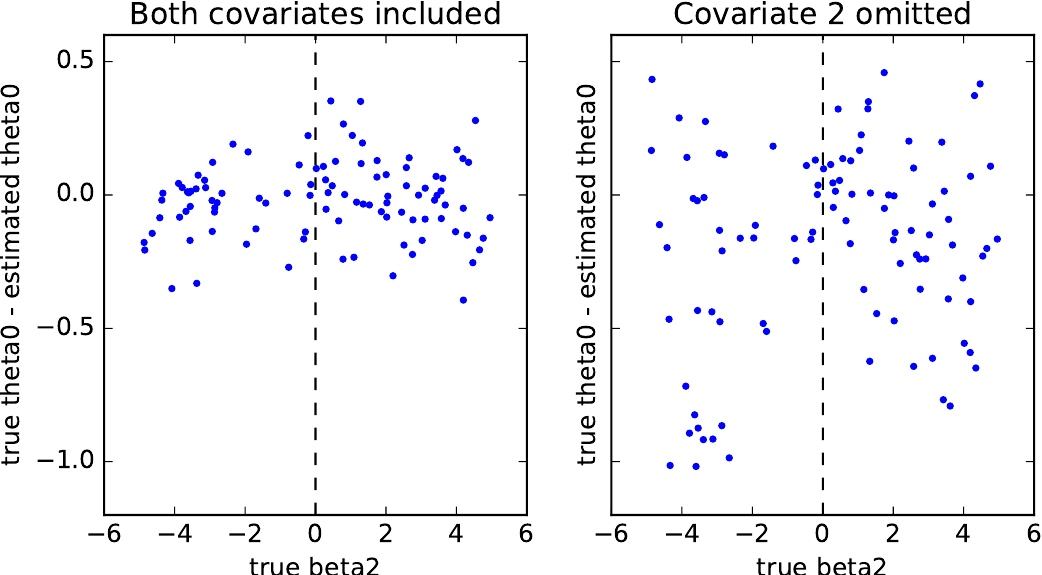}
  \caption{The difference between the true value of \(\theta\) and the estimated
    value, as a function of the coefficient \(\beta_2\), when the two covariates
    are independent. On the left, fits made when \(\beta_2\) is accounted for;
    on the right, when it is not. Notice the odd behavior around \(\beta_2 =
    0\): when the omitted covariate does not matter, \(\theta\) is estimated to
    be close to its true value, but when it has a larger effect, \(\hat \theta\)
    has much higher variance.}
  \label{missingcov-theta0}
\end{figure}

We generated covariates on a grid, drawing the covariate values from a Gaussian
process with squared exponential covariance function to ensure there was some
spatial structure. We first ran 100 simulations (each with new Gaussian process
draws) of independent covariates, fitting a model with both covariates included
and one with the second covariate omitted, each using the expectation
maximization procedure described in Section~\ref{parameter-inference}.
Simulations were performed with random true parameter values, and these values
were recorded, along with the fits. It is apparent from the results that
estimates of \(\hat \theta\) are affected by the missing covariate:
Fig.~\ref{missingcov-theta0} shows the fits, as a function of the true value of
\(\beta_2\) used in the simulation, and a distinct pattern can be seen when the
second covariate is omitted from the fit, with \(\hat \theta\) having larger
variance for larger values of \(|\beta_2|\). On average, the estimated \(\hat
\theta\) with a missing covariate is larger than the true \(\theta\) by
\(0.18\).

% > d2 = read.csv("out/simulation-missingcov-accuracy.csv")
% > median((d2$true_omega - d2$missing_omega) / d2$true_omega)
% [1] -0.6803649

\begin{figure}
  \centering
  \includegraphics[width=0.7\textwidth]{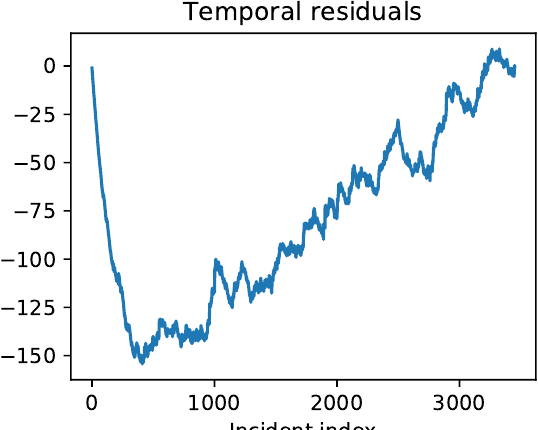}
  \caption{Temporal residual plot for a simulated fit with a missing covariate,
    demonstrating the effect of the overestimated \(\theta\).}
  \label{missingcov-temporal-resids}
\end{figure}

Overestimation of \(\theta\) has other consequences. For example,
Fig.~\ref{missingcov-temporal-resids} shows a temporal residual plot (see
Section~\ref{residual-maps}) for a fit to a simulated dataset with an omitted
covariate. An obvious calibration problem is present: by the time the 500th
event occurred, the conditional intensity function predicted 150 fewer events
than occurred. Near \(t = 0\), \(\lambda(s, t)\) cannot predict the observed
events because there is little past history of events; near \(t = T\), a long
past history and overestimated \(\theta\) causes \(\lambda(s, t)\) to
overestimate the intensity and ``catch up'' in the cumulative predicted number
of events.

Additionally, the time decay parameter \(\omega\) is also overestimated by 70\%
on average. Together, these biases suggest that the clustering induced by the
unobserved covariate is being accounted for by increasing self-excitation and by
allowing the effects of self-excitation to last longer in the model.

\begin{figure}
  \centering
  \includegraphics[width=0.4\textwidth]{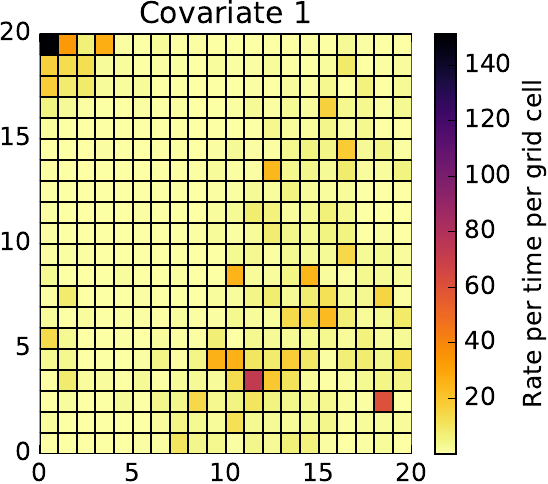}
  \includegraphics[width=0.4\textwidth]{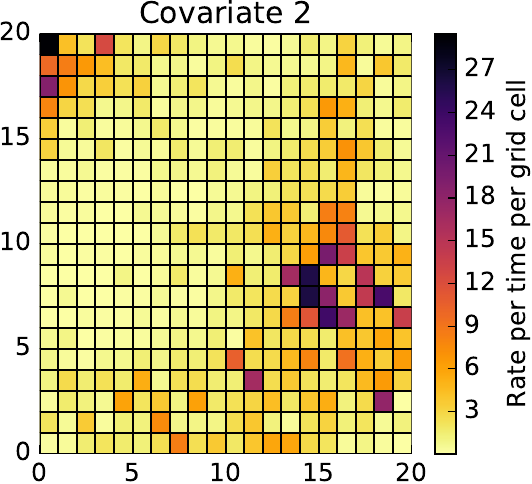}
  \caption{The rate induced (that is, \(\exp(\beta X)\), where \(\beta = 1\) for
    simplicity and \(X\) is the covariate) by two Gaussian process covariates on
    a \(20 \times 20\) grid. The second covariate is dependent upon the first.
    Notice the spatial structure of the Gaussian process.}
  \label{gp-covs}
\end{figure}

\begin{figure}
  \centering
  \includegraphics[width=0.8\textwidth]{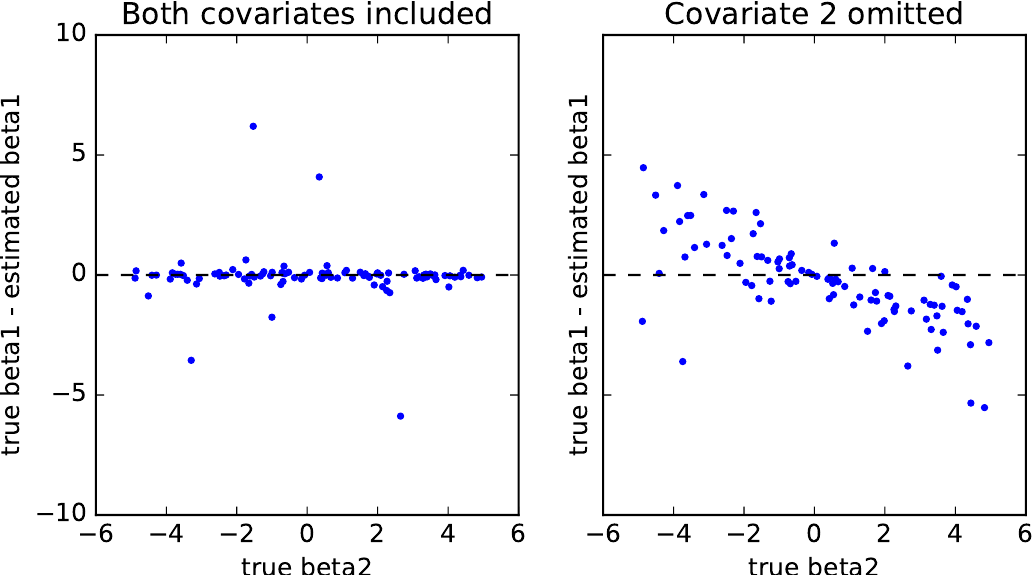}
  \caption{Bias observed in estimated values of \(\beta_1\) when \(\beta_2\) is
    also estimated (left) or is omitted from the fit (right).}
  \label{missingcov-confounded-beta}
\end{figure}

Next, we simulated causally confounded covariates, following the causal model in
Fig.~\ref{confounding-omitted}. Covariate 1 was drawn from a Gaussian process,
as before, and Covariate 2 was defined to be the average of Covariate 1 and a
separate independent Gaussian process. This gave an average correlation of \(r =
0.66\) between the covariates. Sample correlated covariates are shown in
Fig.~\ref{gp-covs}. Data was simulated from these covariates (with random
coefficients) and then models fit with and without Covariate 2 included.
Fig.~\ref{missingcov-confounded-beta} demonstrates the bias in estimates of
\(\beta_1\) that ensues when the effect of \(\beta_2\) is not accounted for,
similar to the biases that can occur in ordinary linear regression when
covariates are confounded. The confounding also affects \(\hat \theta\) and
\(\hat \omega\) in a similar way as in the previous simulation, with bias as
\(|\beta_2|\) increases.

Together, these simulations demonstrate two important caveats of self-exciting
point process models:

\begin{enumerate}
\item Omitted spatial covariates, whether or not they are confounded with
  observed covariates, can bias estimates of the self-excitation parameter
  \(\theta\), making it seem as though events are more likely to trigger
  offspring events.
\item Omitted spatial covariates can also bias estimates of the temporal decay
  parameter \(\omega\), making it seem as though self-excitation or near-repeat
  effects occur over a longer timescale than they really do.
\item If there is a confounding relationship between covariates, such as that
  shown in Fig.~\ref{confounding-omitted}, unobserved covariates can bias
  estimates of observed covariate effects (\(\hat \beta\)) as well as of
  self-excitation.
\end{enumerate}

The first two points are particularly concerning, since in practical
applications it is unlikely that all covariates could ever be accounted
for---there will always be unmeasured spatial differences in base rates, or
imperfectly measured covariates. This suggests that previous applications of
self-exciting point process models may have overestimated the amount and time
scale of self-excitation in the process, unless their background estimator was
able to capture all spatial variation in base rates.

In some cases, the residual analyses introduced in Section~\ref{residual-maps}
may make it possible to detect when there is an important unobserved spatial
covariate. Temporal residual plots like Fig.~\ref{missingcov-temporal-resids}
can suggest the presence of unobserved covariates (or other misspecification),
while residual maps can make systematic deviations from the predicted event rate
visible, and careful examination of the maps may suggest variables that need to
be included. Section~\ref{results} gives examples of this in Pittsburgh crime
data.

General approaches to account for unobserved covariates are more difficult. One
strategy, sometimes used in spatial regressions, is to include a spatial random
effect term intended to account for the unobserved covariates. However, at least
in spatial regression, this method does not achieve its goal: a spatial random
effect can bias coefficients of the observed covariates in arbitrary ways,
particularly if the unobserved covariate is spatially correlated with any of the
observed covariates \citep{Hodges:2010}. Given the causal diagram in
Fig.~\ref{confounding-causal}, it does not seem possible for any one adjustment
to account for an unobserved covariate and give unbiased estimates of the
effects of the other covariates. Users of spatial regression and the
self-exciting point process model introduced here need to be aware of their
limitations in the presence of unobserved confounders, and interpret results
carefully.

\section{Application}
\label{results}

\subsection{Pittsburgh incident data}

To demonstrate the spatio-temporal model of crime proposed here, we will analyze
a database of 205,485 police incident records filed by the Pittsburgh Bureau of
Police (PBP) between June 1, 2011, and June 1, 2016, specifying the time and
type of each incident and the city block on which it occurred. (Privacy
regulations prevent PBP from releasing the exact addresses or coordinates of
crimes, so PBP provides only the coordinates of the block containing the
address.) The records include crimes from very minor incidents (such as 38
violations of Pittsburgh's ordinance against spitting) to violent crimes, such
as homicides and assaults. Only crimes reported to PBP are included, so the
dataset does not include records from the police departments of Pittsburgh's
several major universities, including the University of Pittsburgh, Carnegie
Mellon University, Chatham University, or Carlow University.

Because the database contains only incident reports, offense types are
preliminary. Charges listed in the reports may be downgraded or dropped,
suspects acquitted, or new charges filed. The reports represent only the charges
reported by the initial investigating officers, so they may not correspond with
final FBI Uniform Crime Report data or other sources. While this limits the
accuracy of our data, it is also the only practical approach---final charges may
not be known for months, so predictions based on them would be hopelessly out of
date.

Rather than dealing with the numerous sections and subsections of the
Pennsylvania Criminal Code represented in the incident data, we used the FBI
Uniform Crime Report hierarchy, which splits incident types into a common
hierarchy comparable across states and jurisdictions. Among so-called ``part I''
crimes, homicide, assault, and rape are at the top of the hierarchy, followed by
other crimes like theft, burglary, and so on. If an incident involves two
distinct types of crime (e.g.\ a burglary involving an assault on a homeowner),
we use the type higher in the hierarchy, following the FBI's ``Hierarchy Rule''
\citep{UCR}. The hierarchy of offenses is shown in Table~\ref{ucr-hierarchy}. In
our analysis we focused on crimes in these categories, though other ``part II''
crimes, such as simple assault and vandalism, are also available in the dataset,
along with every other offense type recorded by the Pittsburgh Bureau of Police.
Note that arson, typically hierarchy level 8, was miscoded in the data available
to us, though arson was not used in any of our analyses.

\begin{table}
  \caption{\label{ucr-hierarchy}The part I crime hierarchy}
  \centering
  \begin{tabular}{r l r}\toprule
    \textbf{Hierarchy} & \textbf{Crime} & \textbf{Count}\\ \midrule
    1 & Homicide & 300\\
    2 & Forcible rape & 893 \\
    3 & Robbery & 5884 \\
    4 & Aggravated assault & 5900 \\
    5 & Burglary & 11943 \\
    6 & Larceny/theft & 37487 \\
    7 & Motor vehicle theft & 3892 \\
    8 & Arson & 0 \\
    \bottomrule
  \end{tabular}
\end{table}

We also collected, from city and Census Bureau data, various spatial covariates
for each Census block, including
\begin{itemize}
\item The fraction of residents who are male from age 18--24
\item The fraction of residents who are black
\item The fraction of homes that are occupied by their owners, rather than
  rented
\item The total population
\item Population density (per square meter)
\item The fraction of residents who are black or Hispanic.
\end{itemize}

Some city blocks have no population (e.g.\ in commercial areas with no
residents), so an additional dummy variable was used to record whether each
block had a population. In all models that follow, population-based covariates
only enter the models when the block has a nonzero population.

These covariates will be used to demonstrate the model's ability to account for
spatial factors that attract crime. They are not intended to be a comprehensive
list of all possible risk factors, and undoubtedly there are other relevant
covariates; systematic identification and evaluation of relevant spatial
features is out of the scope of this work.

\subsection{Burglary analysis}
\label{burglary-analysis}

\begin{figure}
  \centering
  \includegraphics[width=0.7\textwidth]{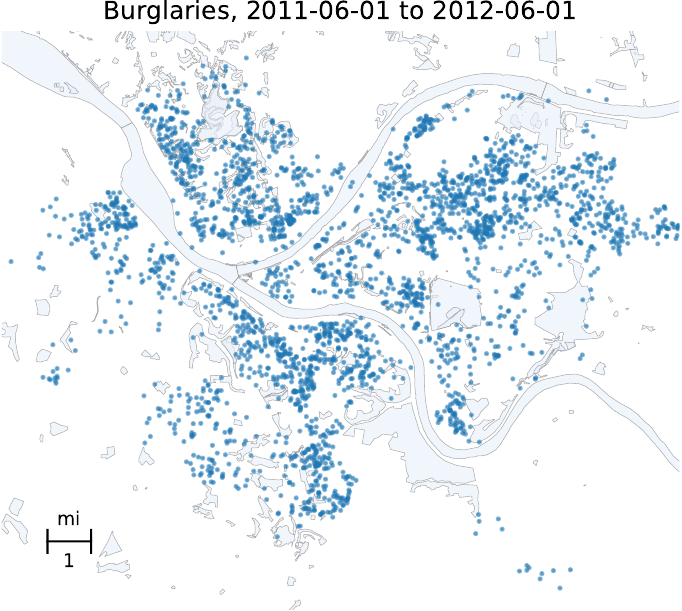}
  \caption{Locations of 2892 burglaries recorded in Pittsburgh between June 1,
    2011 and June 1, 2012}
  \label{burglary-1year-map}
\end{figure}

Selecting only the first year of burglary data, containing 2892 burglaries, we
fit two models, one using only population density as a covariate and the other
using additional covariates. The burglaries are mapped in
Fig.~\ref{burglary-1year-map}, showing spatial structure in the locations of
burglary hotspots across the city. The model fits are shown in
Table~\ref{burglary-1year} and Table~\ref{burglary-1year-covariates}.
Asymptotically normal 95\% confidence intervals based on Rathbun's covariance
estimator are also shown for each parameter. The additional covariates improve
the model AIC from 179750 to 179319, an improvement of about 431 units. Notice
the relative consistency of the self-excitation parameters \(\hat \omega\) and
\(\hat \sigma^2\) between fits, and that, as expected from the discussion in
Section~\ref{omitted-covariates}, \(\hat \theta\) decreases when additional
covariates are added.

Interpretation of the model with all covariates
(Table~\ref{burglary-1year-covariates}) is straightforward. Each burglary
stimulates or predicts, on average, \(\hat \theta = 0.59\) further burglaries,
distributed with a spatial bandwidth of \(\hat \sigma \approx 468\) feet at a
rate exponentially decaying in time with parameter \(\hat \omega \approx 47\)
days. High population densities predict higher risks of burglary, as there are
more residences to burgle; similarly, blocks with populations greater than zero
(a proxy for residential vs.\ commercial blocks) have a higher burglary rate.
The remaining covariates enter the model when blocks have a population greater
than zero. Higher proportions of young men indicate a lower burglary risk,
though the confidence interval for this effect overlaps zero. Home ownership,
rather than renting, has a negative effect, while a higher fraction of black
residents is correlated with higher burglary rates; these last two factors are
likely confounded with poverty and socioeconomic status, which have strong
relationships with crime but are not included in this model.

\begin{table}
  \caption{\label{burglary-1year}Predicting burglary using population density}
  \centering
  \begin{tabular}{l r r}\toprule
    \textbf{Parameter} & \textbf{Value} & \textbf{CI} \\ \midrule
    $\theta$ & 0.764 & [0.717, 0.811] \\
    $\omega$ & $4.511 \times 10^6$ (52.21 days) & [47.04, 57.39] \\
    $\sigma^2$ & $2.664 \times 10^5$ (516.1 feet) & [487.2, 543.5] \\
    $\beta_0$ (intercept) & -31.63 & [-31.50, -31.76] \\
    $\beta_1$ (pop / $\text{m}^2$) & 31.66 & [8.91, 54.4] \\\midrule
    AIC & 179750 & \\
    \bottomrule
  \end{tabular}
\end{table}

\begin{table}
  \caption{\label{burglary-1year-covariates}Predicting burglary using additional
    covariates}
  \centering
  \begin{tabular}{l r r}\toprule
    \textbf{Parameter} & \textbf{Value} & \textbf{CI} \\ \midrule
    $\theta$ & 0.589 & [0.544, 0.635] \\
    $\omega$ & $4.061 \times 10^6$ (47.00 days) & [41.97, 52.04] \\
    $\sigma^2$ & $2.194 \times 10^5$ (468.4 feet) & [439.0, 496.1] \\
    $\beta_0$ (intercept) & -33.15 & [-33.53, -32.78] \\
    $\beta_1$ (pop / $\text{m}^2$) & 25.50 & [6.13, 44.86] \\
    $\beta_2$ (block populated?) & 2.49 & [2.05, 2.92] \\
    $\beta_3$ (frac.\ male 18--24) & -0.69 & [-1.74, 0.36] \\
    $\beta_4$ (frac.\ black) & 0.75 & [0.55, 0.95] \\
    $\beta_5$ (frac.\ homes owned) & -1.14 & [-1.40, -0.88] \\\midrule
    AIC & 179319 & \\
    \bottomrule
  \end{tabular}
\end{table}

\begin{figure}
  \centering
  \includegraphics[width=0.8\textwidth]{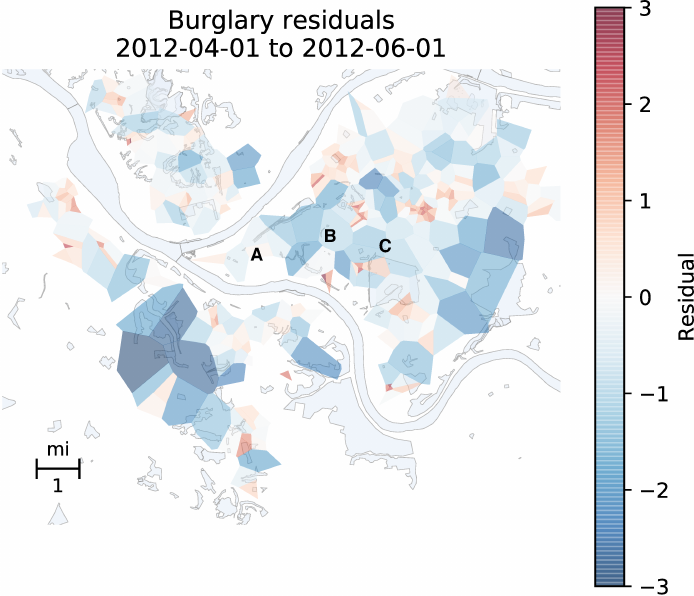}
  \caption{Residual map from the fit shown in Table~\ref{burglary-1year}, over
    two months of burglaries. A: downtown, B: University of Pittsburgh, C:
    Carnegie Mellon University.}
  \label{burglary-resids}
\end{figure}

For a larger view of Pittsburgh, Fig.~\ref{burglary-resids} shows an overall
residual map of Pittsburgh over two months. Several trends appear, suggesting
inadequacies in the available covariates and the presence of boundary effects:
commercial areas such as downtown (at the confluence of the two rivers) have
fewer burglaries than predicted, and the presence of the University of
Pittsburgh and Carnegie Mellon University also results in negative residuals, as
each has its own police department whose records are not included in our
dataset. Note that, as discussed in Section~\ref{residual-maps}, negative (blue)
residuals visually dominate, because areas with lower-than-expected crime hence
have larger Voronoi cells; also note the presence of several clusters of small
cells with large positive residuals, at the locations of temporary burglary
hotspots.

To demonstrate leading indicators, we fit an additional model containing the
same set of covariates but also two leading indicators, larceny/theft and motor
vehicle theft (hierarchy levels 6 and 7). The fit is shown in
Table~\ref{burglary-1year-covariates-leading}, and shows that motor vehicle
theft in particular seems predictive of burglary, with a coefficient of \(\hat
\theta_2 = 0.1167\). The AIC of this model further improved to 179201, by 118
units.

\begin{table}
  \caption{\label{burglary-1year-covariates-leading}Predicting burglary with
    leading indicators}
  \centering
  \begin{tabular}{l r r}\toprule
    \textbf{Parameter} & \textbf{Value} & \textbf{CI} \\ \midrule
    $\theta_0$ (self-excitation) & 0.4480 & [0.404, 0.492] \\
    $\theta_1$ (larceny/theft) & 0.0632 & [0.049, 0.078] \\
    $\theta_2$ (motor vehicle theft) & 0.1167 & [0.037, 0.197] \\
    $\omega$ & $3.551 \times 10^6$ (41.10 days) & [36.5, 45.8] days \\
    $\sigma^2$ & $1.619 \times 10^5$ (402.3 feet) & [376, 427] feet \\
    $\beta_0$ (intercept) & -33.90 & [-34.47, -33.33] \\
    $\beta_1$ (pop / $\text{m}^2$) & 25.19 & [2.48, 47.9] \\
    $\beta_2$ (block populated?) & 3.00 &  [2.37, 3.63] \\
    $\beta_3$ (frac.\ male 18--24) & -0.85 & [-2.14, 0.43] \\
    $\beta_4$ (frac.\ black) & 0.94 & [0.72, 1.15] \\
    $\beta_5$ (frac.\ homes owned) & -1.00 & [-1.30, -0.71] \\\midrule
    AIC & 179201 & \\
    \bottomrule
  \end{tabular}
\end{table}

Temporal calibration plots for these models show patterns similar to
Fig.~\ref{missingcov-temporal-resids}, suggesting, as discussed in
Section~\ref{omitted-covariates}, that there is additional spatial
heterogeneity in crime rates which is not accounted for by the available
covariates in these models, and hence that the self-excitation parameters may be
overestimates. Further research is necessary to identify relevant covariates and
prepare higher-resolution covariate datasets to adequately model crime.

\section{Conclusions}

% TODO there have been applications in various areas, and here we build on these
% to introduce a model for crime

Self-exciting point processes have been used for a wide range of applications,
from epidemiology to seismology, and we have built on this work to introduce an
improved model for crime, extending previous crime models by incorporating
spatial covariate information and providing parameter inference tools to aid
understanding of the patterns of crime. Though self-exciting point process
models are more complex than ordinary spatial regression, making analysis more
difficult for users used to the wealth of tools available for regression, we
have helped bridge this gap through interpretable residual diagnostic tools
(adapted from related models) and through scoring methods for comparing the
predictive performance of models, neither of which has previously been used with
any crime hotspot analysis tool.

A contribution of our work is a demonstration, both theoretically and through
simulations, that methods that focus purely on the spatial or temporal aspects
of crime are generally confounded and can produce misleading results, requiring
a method that accounts for both aspects simultaneously. This calls into doubt
previous results on the connection of spatial features to crime, and the problem
generalizes to self-exciting processes outside of crime, such as models of
infectious disease and earthquakes. Extensive simulations characterize our
model's reaction to misspecification and omitted covariates, both likely
problems to experience in real-world data.

Together, the tools and simulations presented in this paper provide a single
comprehensive package of modeling, diagnostic, and inference tools for
self-exciting point processes, which have not previously been assembled in one
place. The model and tools will enable new criminological research, revealing
patterns of crime, allowing tests of theories about the origin and dynamics of
crime, and contributing to improved policing strategies. Additional research
will more extensively explore the Pittsburgh crime dataset, along with other
cities, and additional covariates and types of crime. Further, the tools and
results described here apply beyond the analysis of crime, to any
spatio-temporal process with self-excitation.

\section*{Acknowledgments}

Thanks to Daniel S.\ Nagin for criminological advice, and to Evan Liebowitz for
compiling the Pittsburgh spatial covariate data. We thank the anonymous referees
for suggestions which substantially improved the manuscript.

This work was supported by Award No.\ 2016-R2-CX-0021, awarded by the National
Institute of Justice, Office of Justice Programs, U.S.\ Department of Justice.
The opinions, findings, and conclusions or recommendations expressed in this
publication are those of the authors and do not necessarily reflect those of the
Department of Justice.

{\small
\bibliography{references}
}

\end{document}